\documentclass[conference]{IEEEtran}
\IEEEoverridecommandlockouts
\usepackage{cite}
\usepackage{amsmath,amssymb,amsfonts}
\usepackage{algorithmic}
\usepackage{graphicx}
\usepackage[section]{placeins}
\usepackage{caption}
\usepackage{textcomp}
\usepackage{xcolor}
\usepackage{float}
\usepackage{wrapfig}
\renewcommand{\thefigure}{\arabic{figure}}
\usepackage[utf8]{inputenc}

\title{\LARGE \bf
AI-Driven Chatbot for Intrusion Detection in Edge Networks: Enhancing Cybersecurity with Ethical User Consent
}

\author{
    \IEEEauthorblockN{Mugheez Asif\textsuperscript{1},
                      Abdul Manan\textsuperscript{1},
                      Abdul Moiz ur Rehman\textsuperscript{1},
                      Mamoona Naveed Asghar\textsuperscript{2},
                      Muhammad Umair\textsuperscript{1}{*}}
    \\
    \IEEEauthorblockA{\textsuperscript{1}Faculty of Information Technology and Computer Science,
    University of Central Punjab, Lahore, Pakistan\\
    \IEEEauthorblockA{\textsuperscript{2}School of Computer Science,
    University of Galway, Galway, Ireland}
    \\
    {*}muhammad.umair@ucp.edu.pk}
}
\begin{document}
\maketitle{}
\thispagestyle{empty}
\pagestyle{empty}

\begin{abstract}

In today’s contemporary digital landscape, chatbots have become indispensable tools across various sectors, streamlining customer service, providing personal assistance, automating routine tasks, and offering health advice. However, their potential remains underexplored in the realm of network security, particularly for intrusion detection. To bridge this gap, we propose an architecture chatbot specifically designed to enhance security within edge networks specifically for intrusion detection. Leveraging advanced machine learning algorithms, this chatbot will monitor network traffic to identify and mitigate potential intrusions. By securing the network environment using an edge network managed by a Raspberry Pi module and ensuring ethical user consent promoting transparency and trust, this innovative solution aims to safeguard sensitive data and maintain a secure workplace, thereby addressing the growing need for robust network security measures in the digital age.

\end{abstract}

\section{INTRODUCTION}

As the digital landscape continues to evolve, the significance of cybersecurity has grown exponentially. Cybersecurity involves protecting computers, networks, and data from unauthorized access, aiming to shield individuals, businesses, and organizations from the detrimental effects of cyber-attacks. The reliance on technology has heightened the risks associated with financial losses, identity theft, and privacy violations, making robust cybersecurity measures more critical than ever. Traditionally, cybersecurity experts have been responsible for identifying and mitigating cyber threats using Security Information and Event Management (SIEM) tools. Such tools use the user’s information without ethical consent from the user. Simultaneously, chatbots are becoming increasingly popular across various domains due to their ability to streamline processes and reduce human workload. These AI-driven tools are widely used for customer service, personal assistance, routine task automation, and providing health advice. Notable examples include Babylon Health, ChatGPT, Aivo, and Mitsuku, which have demonstrated the efficacy of AI in enhancing service delivery and user experience. This growing popularity of chatbots underscores their potential to be leveraged in new and critical applications, such as cybersecurity.\\
In response to the growing challenges and opportunities in cybersecurity, we propose an AI-powered chatbot specifically designed for intrusion detection in network security. Implemented using a Raspberry Pi, we have created a virtual network where, upon connection, the user first verifies their identity using OTP (One time passcode) system and then provided with an ethical consent to the monitoring process. The chatbot excels at intrusion detection, continuously analyzing user activities to detect and mitigate threats. This innovative solution demonstrates the potential of integrating advanced AI technologies with chatbot systems to significantly enhance cybersecurity measures, particularly within edge networks. The intuitive design allows users to easily interact with the system, provide necessary information, and understand the network's security policies. These promising outcomes highlight the system's capability to significantly improve intrusion detection measures while maintaining transparency by taking user consent for monitoring on edge devices. Consequently, helping to reduce manpower for cybersecurity.
\setlength{\parskip}{0pt}
\section{RELATED WORK}
The following section provides an overview of Security Information and Event Management (SIEM) systems and their trends. It also reviewed analytical studies on the general chatbot trends and their application towards network monitoring, identification of malicious activities by the edge networks and Intrusion Detection System (IDS).
\setlength{\parskip}{0pt}
\subsection{Security Information and Event Management (SIEM) Tools}

SIEM tools are the most important part of IT security when it comes to managing critical infrastructures security. A systematic review of current popular SIEM systems, analyzed their capabilities, and provided insights on improvement for the next generation \cite{C1}. It focused on the ability of SIEM tools in highlighting the risks and minimizing the costs and time towards incident investigation. However, present SIEM tools show scalability and visualization challenges and lack the tools to aid decision making concerning the large volumes of events. Moreover, another research work focused on the live analysis technique for Intrusion Detection Systems (IDS) by using Machine learning along with SIEM which improves the real time cyber-attack detection and monitoring \cite{C2}. They conducted performance testing to compare the resources used by various tools and found that the Splunk tool was the most resource demanding among others while the Zeek tool was the least resource demanding out of all the tools analyzed in their study. However, it should be noted that these technologies secretly collect information about users which raises ethical concerns regarding user privacy rights.

\subsection{Chatbot Trends}
Chatbots have become more popular over the years in multiple fields such as education, banking, health care and customer relations etc. In \cite{C3}, a study on the use of chatbots in education, observing that the majority of research utilize quantitative approaches such as ANOVA and correlation analysis, with an increased focus on directed learning. They highlight an absence of different learning methodologies and limited use of chatbots in K-12 education and other areas except for language learning. In the same regard, it was found that chatbots increase motivation and satisfaction in university EFL settings \cite{C4}. Similarly, another study discussed the future of intelligent chatbots in customer service and argued the importance of relating research and development to real-life applications. They point out that present chatbot functions are still in their early phases and they emphasize that it is crucial to have objective standards for the usage of these tools in real life \cite{C5}. Factors influencing chatbot adoption among university students, verifying sections in the Technology Adoption Model and noting that there is a need for cross-country study to increase the generalizability \cite{C6}. Moreover in \cite{C7}, they explored the factors that lead to user’s loyalty to ChatGPT, and they found out that information quality and perceived effectiveness significantly affects the user satisfaction and loyalty. They also highlight that ethical issues negatively influence this relationship, emphasizing that they should be considered when implementing AI chatbots.\\
It was highlighted that although the use of AI-based chatbots in customer service is efficient in terms of cost and time, the efficiency negatively impacted the perception of user satisfaction and created user skepticism \cite{C8}. They also recommend future research on the new design paradigms that can help enhance users’ level of compliance and interaction satisfaction. Furthermore in \cite{C9}, the use of AI chatbots in education, finding major benefits for educators as well as raising concerns about dependability, accuracy, and ethical issues raised where more well-controlled, long-term research is needed. Additionally, chatbot usage intentions in veterinary consultations, highlighting the need for additional study on long-term adoption and user satisfaction across diverse demographics \cite{C10}. A study described the factors explaining the use of messenger chatbot in e-commerce and found that anthropomorphic design and privacy protection are critical for high consumer trust and happiness \cite{C11}. They emphasized the research's importance on how to improve the accuracy of customer profiles and how to manage the privacy aspects efficiently. Lastly in \cite{C12}, the rapid adoption of AI chatbots during the COVID-19 epidemic, emphasizing the importance of research on business preparedness, deployment methods, and ongoing AI training to suit changing consumer demands.
The template is used to format your paper and style the text. All margins, column widths, line spaces, and text fonts are prescribed; please do not alter them. You may note peculiarities. For example, the head margin in this template measures proportionately more than is customary. This measurement and others are deliberate, using specifications that anticipate your paper as one part of the entire proceedings, and not as an independent document. Please do not revise any of the current designations
\setlength{\parskip}{0pt}
\subsection{Chatbots in Network Monitoring }
Chatbots are increasingly used to monitor various systems and provide real-time updates. In \cite{C13}, a new approach to monitor the hydroponic system using a chatbot on Telegram. They have described through their work how IoT sensors can be connected to Firebase as a back-end database for efficient transmission of data in real-time with an average access time of 1. 44 seconds. In the field of mental health, an AI chatbot for the remote monitoring of mental health which was based on the methods of the Behavioral Activation (BA) treatment was developed \cite{C14}. Moreover, the entire risks and threats associated with chatbots were examined together with the focus on private life and data protection \cite{C15}. Their work covers the aspect of proper and secure management, and processing of data and information where there is no prescribed set of rules as the GDPR. The research related to the innovative use of AI-based chatbots for the intervention of behavior change was also seen \cite{C16}. The study also highlights the importance of collecting data on real time interactions for personal tailored health therapies, and the fact that prior studies are moderately to highly at risk for internal validity. They suggest that future research should employ more robust methods and provide clearer descriptions of AI tasks to ensure that health chatbots are effective and can be deployed in as many contexts as possible. Moreover, chatbot’s capacity to learn and comply with corporate processes by examining over 500,000 customer support interactions \cite{C17}. Their technique measured chatbots capacity to follow normative process models. They notice the need to establish common process mining metrics for evaluating chatbot effectiveness and agreed that such functional studies should be further investigated by integrating process compliance assessment and current NLP-based methodologies for better performance, particularly in business settings.
\subsection{Detecting Malicious Activities on Edge Network and Intrusion Detection System (IDS)}
Identification of malicious activities via edge networks and
Intrusion Detection Systems (IDS) are difficult as cyber-attacks are becoming more complex day by day. It was highlighted that blockchain has the capability of safeguarding the IoMT devices in healthcare and ensuring secure records transfer. Their study shows the need for tailoring blockchain-based security mechanisms specifically designed for IoMT edge networks \cite{C18}. Additionally, pAElla – an edge computing-based AI framework for proactively detecting malware from data centers and supercomputers in real-time was introduced \cite{C19}. According to their work, they achieved pretty high accuracy, nearly equal to 1 with F1-score but suggest future studies for further enhancement of the method to verify its applicability in large-scale productions and real-time. Nonetheless, a cybersecurity framework for identifying malicious edge devices in fog computing environments was proposed \cite{C20}. Their approach was tested with genuine assaults and demonstrated efficacy in lowering the false alert rate of IDS. In \cite{C21}, a model for a secure network to mitigate bot attacks in edge-based IIoT applications to protect from both known and unknown threats was proposed, which had positive results. They emphasize the importance of further optimizing machine learning and deep learning algorithms for practical applications. Proposal of an edge-based blockchain-enabled anomaly detection solution for defending IoT networks against insider threats has also been done \cite{C22}. Their approach includes applying sequence-based detection method along with decentralization of a block chain that has potential to enhance the data integrity and security. An edge-based system IOT-KEEPER was provided for the detection of malicious IoT network activities with the help of internet traffic analysis. In this paper it was testbed on real-world and IOT-KEEPER achieved relatively high accuracy and very low false positive rates \cite{C23}. Moreover, a proposal for the combination of the edge computing technique with deep learning for the identification of IIoT malware in smart factories was recommended \cite{C24}.\\
Although their technology performed well in controlled studies, the study emphasizes the importance of real-world validation to assure scalability and efficacy in a variety of industrial contexts. However, a study presented a vulnerability assessment technique for 5G core security in edge networks. They underline the need of combining this method with autonomous intrusion response systems to deploy defenses against these cyberattacks \cite{C25}. Additionally, an IDS was developed for wireless control networks to detect malicious behavior, proving its capacity to decrease monitoring overhead by viewing only a subset of nodes \cite{C26}. Moreover, the DTARS framework for threat monitoring in multi-access edge computing systems was introduced in research \cite{C27}. When evaluated in multiple MEC case studies, the architectural approach was proved to be efficient in addressing the DDoS and botnet attacks. Further, they explored the applications of machine learning methods to perform the task of identifying the undesirable traffic in the context of cloud computing \cite{C28}. Their work presented new forms of the additional column that can enhance the quality of identification of machine learning models; expressively highlighted the issue of feature selection and feature construction. In a study the NSL-KDD dataset was used required to compare the results obtained from several classification methods in order to detect the abnormalities in the network traffic \cite{C29}. From the observation, they were able to identify how the protocols in the network were related with the types of attacks which in return helped in enhancing the appreciation of network security. Security using AI models in different security situations such as network, mobile and IOT security was explored \cite{C30}. In the given paper, the issues connected with application of AI in cybersecurity systems and the ways that can be used for the introduction of the solution to the existing systems were described. Additionally, a new IDS model called Deep Q-Learning MAFSIDS is proposed to cope with the problems brought by massive imbalanced data sets and emerging network attacks \cite{C31}. Through a deeper analysis, the study concluded the following: The application of a feature selection technique and the deep reinforcement learning assault detection module raised the detection accuracy prominently. Lastly, a detailed study on Intrusion Detection Systems (IDS) was also arranged which pointed out that IDS has a significant demand in current days due to concern with strong network traffic and security threats \cite{C32}. In their research, they put focus on breakthrough issues that include the need to address multiple intrusion kinds as well as the hefty processing that is always demanded.\\
As shown in several fields of application such as education, health, customer relations, support with mental health and in securing network, chatbots provide valuable and effective assistance. However, this study found a gap in ethical usage of chatbots in gaining consent from the users and it raised issues of privacy and data protection. Likewise, monitoring is used in a variety of chatbots, there is a complete lack of references to their use in network security or intrusion detection. These problems are solved in the proposed solution by using the possibility of the application of chatbot technologies to collect the network packets in real-time, the use of artificial intelligence for intrusion detection, and constant observing of the users’ activity. This approach not only meets a significant gap in present day technologies but also sets up a practice that is much more moral and productive in meeting the goal of protecting digital structures, making it the new norm in cybersecurity.

\section{METHODOLOGY}
The architecture is sound and offers protection, firm and smooth networking of computers in an organization. Each of them is crucial in keeping the connections secure, safe and perceptive through the straightforward use of technologies and Artificial Intelligence to deliver an optimum safe network environment. The architecture is divided into four major components: \\
1. Virtual Network \\
2. Chatbot \\
3. Authentication and Storage \\
4. Intrusion detection\\
The proposed architecture of the chatbot is given in the figure 1.
\setlength{\parskip}{0pt}
\begin{figure*}[t]
    \centering
    \includegraphics[width=\textwidth]{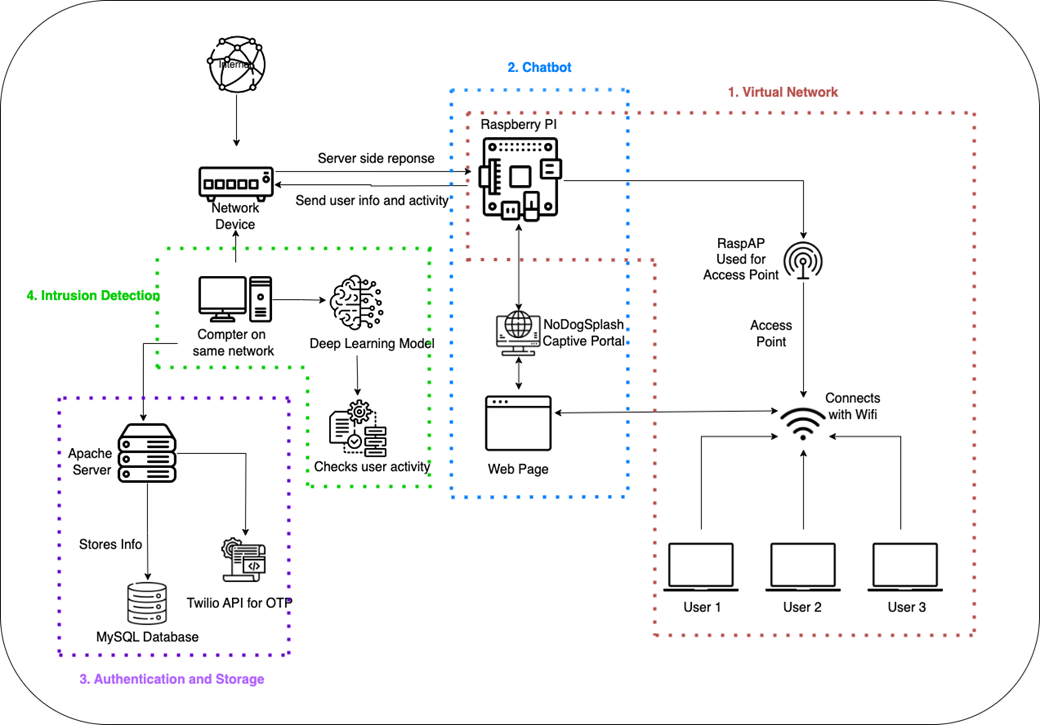} 
    \caption{Proposed Architecture of Intrusion Detection Ethical Chatbot}
    \label{fig:example}
\end{figure*}
\subsection{Virtual Network} In a proactive environment characterized by vigorous interaction in the virtual network, applying a strong and reliable system is critical. This network configuration uses modern technologies in securing the network connections and using the available resources in the network effectively. The infrastructure entails the use of enhanced security measures to ensure that data is not compromised; in this case, WPA2 is adopted for security. One of the key components of this structure is Raspberry Pi Model 5 that performs the function of a virtual network controller. The Raspberry Pi Model 5 comes with better processing power together with better networking capabilities that can be appropriate for handling VNets. It performs the role of being the main connector through which other devices and the network can easily interface.

\subsubsection{RaspAP Configuration}
The virtual network upon which the hosts move about is powered by the Raspberry Pi Model 5 adding the RaspAP package to it. RaspAP is a software package that efficiently turns a Raspberry Pi computer into a working wireless router and an access point. It has an understanding and easy-to-use web graphical user interface to facilitate the setting up of the settings of the network as well as its management to a level that even those with little technical knowledge on the matter will understand. It also enables the Raspberry Pi to serve as an access point and is connected with other networking hardware for instance a router through an Ethernet cable. According to the adjustments made to RaspAP, we have made the network open, thus, connecting end devices is quite easy. This configuration makes it possible for the Raspberry Pi to handle many connections at the same time, this means that the network interface is stable. Configuration of DNS and of firewalling rules among the rest of the network settings is easily possible with the help of the friendly user interface that is shipped with RaspAP package. Besides, response time, traffic, and current connected users and overall network health are also displayed, which is important for proper work. RaspAP was installed the package to Raspberry Pi Model 5, and configured as follows: Using the easy-to-use HTML graphic user interface, we then set up the network name, the security key password, and the DHCP server for the creation of a open network access point. Moreover, configuring the firewall and DNS to suit the network sensitivity provided a strategy of improving network performance and security. 
\subsubsection{WiFi Security}
For security purposes, it has been configured to use WPA2 encryption for the Access Point of the wireless network. WPA2 stands for Wi-Fi Protected Access 2 and is a security standard which aims at offering data protection alongside the task of network authentication. This makes it possible that only the allowed people can be able to connect to the network to secure the important information that may be in the network from being accessed by unauthorized persons. Also from the same package, configuration of guest networks is possible whereby the guest network can only access a restricted set of the primary LAN as a security boost.
\subsubsection{DHCP Management}
As is seen in our virtual network environment, DHCP (Dynamic Host Configuration Protocol) appears to be rather significant. DHCP is a communication protocol used in managing a network to enable the devices on an IP network to access the utility services such as DNS, NTP and any other communication protocol supported by UDP or TCP. RaspAP package has the DHCP functionality, which helps in the proper management of the IP address of the connected devices. This protocol automatically allocates IP addresses to computers and other devices on the network to allow for proper NF and efficient use of resources. DHCP is helpful in automatically assigning and managing the IP addresses associated with a certain network, and this lessens the possibility of connection issues in setting up the network. Such automation is important so that the network should run efficiently, particularly with respect to several concurrent linked clients. However, DHCP employs functions like IP address leasing whereby IP addresses are given to devices on temporary basis, hence can be recycled. Such efficient IP address management allows avoiding situations where different networks use addresses that are identical; it also allows the proper usage of IP addresses. Also, DHCP can convey other information that also enables the network’s configuration including the default router and the DNS server. \\
In summary, combining the functions of RaspAP and DHCP within the proposed architecture of a virtual network guarantees its functionality, adaptability, reliability, and security in the context of today’s networks. This configuration also stabilizes and makes the network operation easy to manage and increases user satisfaction because of their reliable connections.
\setlength{\parskip}{0pt}
\subsection{Chatbot Architecture}
The role of the Chatbot is to offer an access point of the network employing more developed technologies such as Raspberry Pi technology, an OTP protocol for authentication and AI technology for monitoring. This section outlines the procedures and parts that make up the procedure of this system's introduction.
\subsubsection{User Authentication}
Works by automatically opening a web page when a user – device connects to the network and initiating a chat. The web page discussed invites the user to register and requires the input of the mobile phone number. After that, through the Twilio API implemented on the server side, an OTP (One Time Password) is sent to the submitted phone number for confirmation. The OTP is then sent to the user's mobile device and he or she enters the OTP on the web page. The OTP verification leads the web page to the following steps where the network policies will be displayed. These are the regulations that define the network's usage and the ethical monitoring methods. To proceed, the user must read and agree to these policies. If the login is successful, then the system allows the user’s network access and records the event for documentation and audit.
\subsubsection{OTP Verification and Network Policies}Regarding the issue of users not having internet on the first device to inform and receive the OTP, another device with internet connectivity has been provided. This device contains an Apache server with a PHP script that was created to address OTP generation or verification. When the user wants an OTP to be generated, the Rasberry Pi initiates a request in the form of an API to the Secondary device which contains a server-side script. The script then produces the OTP and sends it back to the Raspberry Pi, hence authenticating the user. Thus, this method also guarantees the functionality of the procedure in delivering and confirming the OTP to the said user even without prior connection with the internet. After the OTP process to authenticate the user’s phone number, the last step is to provide the customer with the consent form of the network policies and the general terms of using the chatbot. These policies must be accepted by the user before they can connect to the internet. For confirmation, the system opens Internet connection and records it to keep the record and for auditing.
\subsubsection{User Information Storage and Network Access}
Once the user submits to the policies and acquires Internet Connectivity the system moves to the monitoring mode. This phase uses a machine learning model to carry out network packets analysis within a real-time context to identify any such abnormities. Note that the model can identify anomalies and possible threats to the network and thus maintains security around the clock. The computer's applications are also constantly using this machine learning model for predictive and proactive threat identification and elimination. By analyzing the characteristics of network traffic, the model can detect and counteract tracks of unlawful activities, thus, the network’s integrity, as well as users’ data, can be saved. 
In conclusion, this architecture makes authentication secure, network policies easily understandable and manage the users properly. It uses technologies like Twilio API especially for OTP and verified codes legal data management techniques, and AI monitoring for the network security boost up.
\setlength{\parskip}{0pt}
\subsection{Authentication and Storage}
In terms of Security, the Chatbot architecture duly implements a strong OTP based Authentication to provide good security to Network. The Raspberry Pi used here with the help of the RaspAP library is set up to act as a hotspot and is connected directly to the internet via an Ethernet cable for security. Users when connecting are welcomed by a Nodogsplash captive portal that asks them to input a mobile phone number. A system Call in API’s form is then made to a second device with internet connectivity that carries an Apache system and PHP script for generating and verifying the OTP. Twilio API sends the OTP to the user, and he inserts it on the web page. The Raspberry Pi then checks the OTP received and if it is correct, it approves the user.\\
After the verification process, users carry out the network policies review and approval to get connected to the internet, and the process is recorded by the system. The process guarantees a safe security check of the network as well as the transparent declaration of the policies within the network and good user management from the advanced technologies in reliable network access.
\setlength{\parskip}{0pt}

\subsection{Intrusion Detection }
Once the user agrees to the policies accepting them and subsequently connects to the internet services, the system moves to the monitoring state. During this phase, the application of machine learning models is to paralyze and assess the network packets for any sign of intrusion. The models always watch for emergencies and threats so that protection of the network is always on. The systemic approach to threats identification and their eradication is the key factor in preserving the networks’ safety. The decision tree and the random forest are selected for intrusion detection because of their compatibility in dealing with both qualitative and numerical data and their feature in recognizing nonlinear relationships, which are the usual scenario with intrusions. It will be important to note that decision trees are unique in that they explain the classes and features of importance, can effectively deal with missing data and work out the computations. Decision tree is used in the random forests that create many decision trees and are applied to handle numerous features and imbalanced data to avoid overfitting and increase algorithm efficiency. This combination ensures a reliable instantiation of real-time intrusion detection since it captures relations amongst the features and consolidates the strong and efficient network protection hence providing safety for the users and activities that are being performed online \cite{C28},\cite{C30}. 

\section{GRAPHICAL USER INTERFACE (GUI)}
The following section provides brief information on the different user interface-related screens necessary in the proposed AI chatbot-based architecture for intrusion detection in edge networks. First, the screens help the users go through the procedures of entering their phone number, entering OTP received, as well as agreeing to the network policies. These steps ensure secure access and user consent, which are crucial for ethical monitoring.\\
\setlength{\parskip}{0pt}
\begin{figure}[h] 
    \centering
    \includegraphics[width=0.8\linewidth]{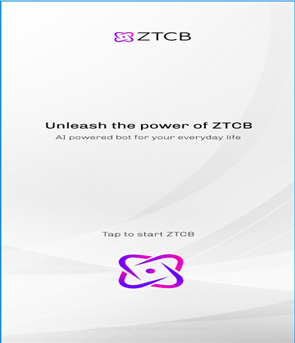}
    \caption{Home screen of ZTCB}
    \label{fig:example}
\end{figure}

The main screen as shown in Figure 2 appears on the device of a user when it establish a connection with the Raspberry Pi. The current version displays ZTCB Logo in the center and underneath the message ‘Tap to start ZTCB’. Following this, the next screen shonw in Figure 3 prompts user to enter his phone number. The user manually type in their phone number to the input field ‘Add Phone number’. Once it is entered, the user is proceeded to the next step of verification by clicking on the ‘Send Code Button’.
\setlength{\parskip}{0pt}
\begin{figure}[h] 
    \centering
    \includegraphics[width=0.8\linewidth]{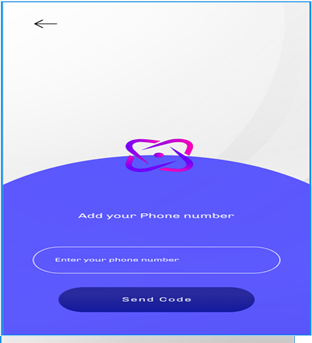}
    \caption{Screen to get user cell number for sending OTP}
    \label{fig:example}
\end{figure}
\begin{figure}[h] 
    \centering
    \includegraphics[width=0.8\linewidth]{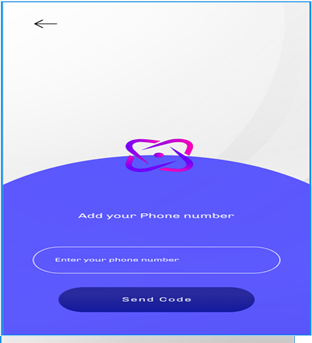}
    \caption{OTP Verification Screen}
    \label{fig:example}
\end{figure}
\\The next screen is shown in Figure 4 requires OTP which is sent to the user’s entered number. Before proceeding, there is a clock that indicates the time and a message to enter the ‘Fetch security code’. If the OTP is not received, then there is an option of “Resend Code” to get the OTP within the specified time. Once OTP verification is done, users are directed to the consent page as in Figure 5. Here user reviews and accepts the network policy. They must check the box and click the connect button to complete the registration and gain network access. Once OTP verification is done, user is free to use network services with backend monitoring through log file creation for intrusion detection.
\setlength{\parskip}{0pt}

\begin{figure}[h] 
    \centering
    \includegraphics[width=0.8\linewidth]{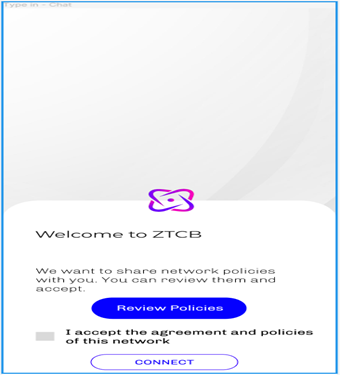}
    \caption{Ethical consent regarding monitoring for intrusion detection}
    \label{fig:example}
\end{figure}
\section{RESULTS}
For training our system, we applied a Decision Tree model and a Random Forest model on the well-known NSL-KDD dataset commonly used for network intrusion detection learning. The results obtained by the decision tree and random forest models are compared to detect intrusion. The evaluation is done from their confusion matrices as outlined below which gives the insight of each model’s accuracy, precision, F1-score and recall. This comparison gives an insight on experts’ evaluation on the potential of each model in perceiving and mitigating network threats. The confusion matrices extracted from our models is indicated below in Figure 6 and Figure 7.
\begin{figure}[h] 
    \centering
    \includegraphics[width=0.8\linewidth]{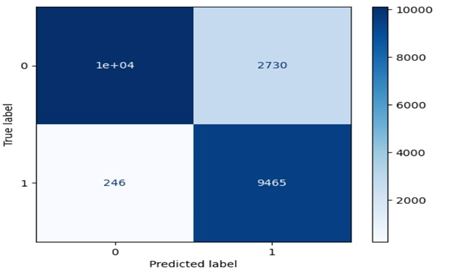}
    \caption{Confusion matrix of Decision tree}
    \label{fig:example}
\end{figure}
\begin{figure}[h] 
    \centering
    \includegraphics[width=0.8\linewidth]{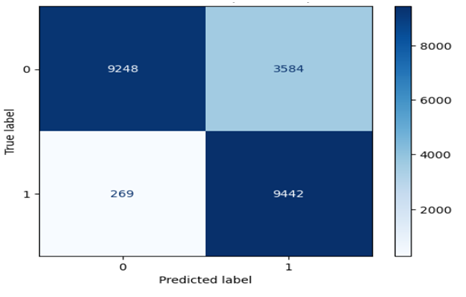}
    \caption{Confusion matrix of Random Forest}
    \label{fig:example}
\end{figure}
\\For the Decision tree model from Figure 6:\\
•	True Positives (TP): 9465, False Positives (FP): 2730, True Negatives (TN): 10000, False Negatives (FN): 246\\
For the Random Forest model from Figure 7:\\
•	True Positives (TP): 9442, False Positives (FP): 3584, True Negatives (TN): 9248, False Negatives (FN): 269\\
From the confusion matrix, we derived several performance metrics to evaluate the effectiveness of our models:
\begin{table}[h!]
\centering
\begin{tabular}{|l|c|c|}
\hline
\textbf{Scores} & \textbf{Decision Tree} & \textbf{Random Forest} \\
\hline
Accuracy      & 86.84\%       & 83.48\%        \\
\hline
Precision     & 77.60\%       & 72.50\%        \\
\hline
Recall        & 97.45\%       & 97.23\%        \\
\hline
F1 Score      & 86.30\%       & 83.85\%        \\
\hline
\end{tabular}
\caption{Performance metrics for Decision Tree and Random Forest}
\label{tab:performance}
\end{table}
The confusion matrix presented in Figure 2 and Figure 3 gives a better understanding of the classifier’s performance, in this case, a Decision Tree and Random Forest with the NSL-KDD dataset. Since both models have a high value of recall, indicating they are both effective at identifying true positive cases of intrusions. However, a tree decision model occurs more efficiently. Particularly, it has larger accuracy and precision compared to the random forest model. Higher accuracy means that the true positive and negative are classified correctly. Higher precision means, if the decision tree model detects an intrusion, then the probability of the model being right is high, thus reducing the number of false positives. \\
This superior performance is important when it comes to an intrusion detection system. High accuracy ensures that the system reliably identifies real threats while minimizing false alarms. Based on these advantages, we decided to incorporate the decision tree model for our intrusion detection system to achieve better threat detection and ensure a stronger security measure for the organization’s network. Despite the high recall showing that the given model successfully detects real attacks, precision is lower here, which in other means some level of false alarms. Altogether, the obtained results seem to be reasonable with the help of F1 score that is used to balance  the model; therefore, it might be useful for network intrusion detection.
\section{Conclusion}

In this paper, we propose a new AI-empowered chatbot architecture for intrusion detection at the edge network level. Our model integrates an advanced AI module with a Raspberry Pi module, enhancing network security robustness and scalability. The system uses OTP-based authentication for user verification and employs a machine learning model to monitor network traffic, identifying and mitigating threats in real-time. Our results demonstrate the system's efficiency in improving network security while maintaining user transparency and consent. Performance metrics for the decision tree model applied to the NSL-KDD dataset are very promising, indicating real-world application potential. Additionally, ethical considerations foster user trust and compliance, addressing privacy and data protection concerns. The architecture's integration of virtual network management, chatbot interaction, and intrusion detection sets a new standard for edge environment cybersecurity. Future work will focus on refining AI algorithms to enhance detection accuracy and expand the system's capability to handle various cyber threats. Our AI-powered chatbot presents a proactive and ethical solution to network security challenges, offering a secure environment for transactions and sensitive data in today's digital world.
\setlength{\parskip}{0pt}

\bibliographystyle{ieeetr}

\bibliography{main}
\end{document}